\documentstyle[12pt]{article}
\def\c60{~C$_{60}$}
\def\be{\begin{equation}}
\def\ee{\end{equation}}
\def\bea{\begin{eqnarray}}
\def\eea{\end{eqnarray}}
\def\Oh{{\hat \Omega}}
\def\half{\frac{1}{2}}
\begin{document}
\baselineskip 18pt
\title{Electron--Vibron Interactions in Charged
Buckminsterfullerene: Pair Energies and Spectra (Part II)}
 \author{Nicola Manini$^a$\thanks{Email: manini@tsmi19.sissa.it},
Erio Tosatti$^{a,b}$\thanks{Email: tosatti@tsmi19.sissa.it},\\
{\it a) International School for Advanced Studies (SISSA),}\\
{\it via Beirut 2-4, I-34013 Trieste, Italy,}\\
{\it b) International Centre for Theoretical Physics (ICTP),}\\
{\it P.O. BOX 586, I-34014 Trieste, Italy,}\\
and\\
Assa Auerbach\thanks{Email: assa@phassa.technion.ac.il}\\
{\it Physics Department, Technion, Haifa, Israel.} }
\maketitle

\begin{abstract}
The ground state energy shifts and excitation spectra of charged
buckminsterfullerene  C$_{60}^{n-}$, $n=1,\ldots 5$ are calculated. The
electron--vibron Hamiltonian of Part I is extended to include  all $A_g$ and
$H_g$ modes with experimentally determined frequencies and theoretically
estimated coupling constants.  Complex splitting patterns of $H_g$ vibrational
levels are found.  Our results are relevant to EPR measurements of spin
splittings in C$_{60}^{2-}$ and C$_{60}^{3-}$ in solution. Spectroscopic
gas-phase experiments will be of interest for further testing of this theory.
As found in Part I,  degeneracies in the electron and vibron Hamiltonians give
rise to a dynamical Jahn Teller effect, and to a considerable enhancement of
the electronic pairing interaction. This helps to overcome repulsive Coulomb
interactions and has important implications for superconductivity in
K$_3$C$_{60}$ and the insulating state in K$_4$C$_{60}$.
\end{abstract}

\section{Introduction}
\label{intro}

This paper continues the investigations of the  electron-vibron interactions
within a fullerene molecular anion, C$_{60}^{n-}$. In Part I (Ref. \cite{I}) we
considered in great detail the idealized case of a single  $H_g$ vibron mode
coupled to the electronic degenerate $t_{1u}$ orbital for $n$=1,..5 electrons.
By solving the problem for strong, intermediate and weak  coupling regimes, we
have shown the existence of Berry phases for odd $n$, and their importance in
determining  ground state energies and degeneracies. The Berry phase effects
are clearest in strong coupling, where the parity  of the pseudorotational
orbital angular momentum $L$,  is related to the electron filling $n$ by
\be
(-1)^{L+n} = 1
\ee
The effects on the energies however are relatively stronger in weak coupling
where quantum corrections enhance pair binding which is a factor 5/2 {\em
larger} than the corresponding  classical  Jahn-Teller (JT) relaxation energy
$E^{JT}$. This  leads to   larger electron-vibron pairing interaction than
previously calculated using Migdal-Eliashberg theory.   Moreover, this
enhancement is of direct importance to C$_{60}^{n-}$, where the electron-vibron
coupling is weak to intermediate. These encouraging results clearly call for a
more realistic study of the full electron-vibron problem of C$_{60}^{n-}$. This
is the purpose of this paper, where we will address both the vibronic spectrum,
and the pairing energies, in quantitative detail.

The full molecular Hamiltonian describes the dynamics of 60 carbon atoms plus
240+$n$ valence electrons. The problem is substantially simplified by assuming
the knowledge of the non interacting spectra for electronic levels
\cite{satpathy,negri,andreoni} and for molecular vibrations, both from theory
\cite{negri,andreoni} and experiment \cite{t1,zhou}. For $n$=0, the electrons
form a closed shell. This allows a Born-Oppenheimer decoupling of the vibron
and electron systems. For $n$=1,...5, however, the extra electrons partly fill
the threefold degenerate Lowest Unoccupied Molecular Orbitals (LUMO) states of
$t_{1u}$ symmetry, which gives rise to a linearly coupled JT system. Other
orbitals, such as the Highest Occupied Molecular Orbitals (HOMO) states of
$h_u$ symmetry, at $\sim-2$eV below the LUMO states, and  the  $t_{2g}$
(LUMO+1) states at $\sim+1$eV,  introduce weaker quadratic electron-vibron
couplings which we presently ignore. Further complications may arise from
anharmonic effects.

Fortunately, neglect of all these higher order effects is expected to be a very
good approximation in C$_{60}^{n-}$, where the $t_{1u}$ level is very well
separated from others, and the C-C bonds are rather stiff and harmonic.
Detailed Hartree-Fock calculations have shown, for example, that the energy
gain by going from $I_h$ to $D_{5d}$, $D_{3d}$, $D_{2h}$ symmetries via static
JT distortions, are in fact identical to within 1\% \cite{koga}. Therefore,
restricting to the $t_{1u}$ orbital, to linear JT coupling and harmonic vibrons
is very well justified in C$_{60}^{n-}$. Neglecting also Coulomb interactions
(they will be discussed separately in Section \ref{MMS}), the full electronic
problem is therefore replaced by a 3$\times$3 matrix linearly coupled to
vibrons.

Thus our Hamiltonian is an extension of the single mode mode solved in full in
Part I. Here we shall include the eight $H_g$ vibron modes of real C$_{60}$,
instead of only one.  Also we shall include two $A_g$ vibrons which also couple
linearly to the LUMO electrons, even though they do not split its degeneracy.
Symmetry prevents all other vibrons different from $A_g$ and $H_g$ to interact
linearly with the $t_{1u}$ orbitals. If we further neglect higher order
interactions, all other vibrons are  decoupled and unaffected by changing the
electronic filling $n$.

Generally speaking, even with these drastic approximations, a realistic
description of the dynamical JT state of a C$_{60}^{n-}$ ion is a more
complicated affair than the single mode treated in Part I. For a general
coupling strength, there is in fact no linear superposition between effects
produced by different $H_g$ vibrons coupled to the {\em same} $t_{1u}$ orbital.
Luckily, however, linear superposition turns out to be valid in the weak
coupling perturbative regime, which, in turn, applies, even if only
approximately, to fullerene.

In this paper we apply perturbation theory to the full electron-vibron problem
of C$_{60}^{n-}$, with $n$=1 to 5 in a $t_{1u}$ orbital, and including all
$A_g$ and $H_g$ modes \cite{ehsymm}. As in Part I, we make the further
approximation of replacing a spherically symmetrical coupling to the true
icosahedral hamiltonian on the C$_{60}$ ball. Analytical expressions correct to
second order in the electron-vibron coupling strengths are found for ground
state and excitation energies, as well as for electron pair binding energies
(''effective Hubbard $U$'s``). Using realistic coupling constants from Local
Density Approximation calculations, we obtain numerical estimates for these
vibron-induced pair energies and find them to be unexpectedly large and
negative ($\approx -0.2$eV) for $n$=1,3,5, and even larger but positive
($\approx 0.4$eV) for $n$=2,4. This finding is discussed in qualitative
connection with superconductivity in A$_3$C$_{60}$ \cite{sch,vzr,commj}, and
with the insulating state of A$_4$C$_{60}$ (A=K,Rb).

The calculated vibron spectrum of C$_{60}^{n-}$ is also presented in detail.
Not\-withstanding the  uncertainty in the physical coupling constants, large
splittings of all $H_g$ modes are predicted.  These are expected to be
observable for example in gas phase C$_{60}^-$ and C$_{60}^{2-}$ ions. This
part is organized as follows: Section \ref{model} defines the multi vibron
model. Section \ref{MMS} describes the perturbative calculation of the
spectrum. Section \ref{spectros} presents the relation of the results to
experimental measurements of vibron spectroscopy of C$_{60}^{n-}$ anions.
Section \ref{gse} discusses the interplay between electron-vibron and Coulomb
interactions for the ground state and pair binding energies. We conclude with a
short summary.

\section{The Hamiltonian}
\label{model}
The single electron LUMO states of \c60 are in a triplet of $t_{1u}$
representation.  The important vibrational modes which couple to this
electronic shell are of two representations: $A_g$ (one dimensional) and $H_g$
(five dimensional). $A_g$, $t_{1u}$ and $H_g$ are the icosahedral group
counterparts of the spherical harmonics  $1$, $\{Y_{1m}\}_{m=-1}^1$, and
$\{Y_{2m}\}_{m=-2}^2 $ respectively. By replacing  the truncated icosahedron
(soccer ball) by a sphere, we ignore lattice corrugation effects which are
expected to be small for the electron-vibron interactions, since they do not
lift the degeneracies of $L=0,1,2$ representations.

The  Hamiltonian includes the terms
\be
H~=H^0 ~+ H^{e-v} ~+ \cdots
\label{2.1}
\ee
where electron-electron interactions, anharmonic interactions between phonon
modes, and anharmonic coupling terms have been neglected. The non interacting
Hamiltonian is
\be
H^0=  \hbar \sum_k \omega_k \sum_{M=-L_k,L_k}
\left(b^\dagger_{ k M } b_{ k M }+\half\right)~+(\epsilon-\mu)\sum_{ms}
c^\dagger_{ms} c_{ms},
\label{2.2}
\ee
$b^\dagger_{kM }$ creates a vibron of mode $k$ and energy $\omega_{k}$  in the
spherical harmonic  state $Y_{L_k M}$, where $L_k$ denotes the angular momentum
of mode $k$, either 0 or 2  according to whether $k$ is an $A_g$ or an $H_g$
representation respectively. $c^\dagger_{ms}$ creates an electron of spin $s$
in an orbital $Y_{1m}$. This Hamiltonian operates on the basis \be
\prod_{KM}|n_{kM}\rangle_v\prod_{ms} |n_{ms}\rangle_e
\label{basis}
\ee
where $|n_{kM}\rangle_v$ ($ |n_{ms}\rangle_e$) is a vibron (electron) Fock
state. By setting $\mu\to \epsilon$ we discard the second term in (\ref{2.2}).

The electron--vibron interaction is local and we assume it to be rotationally
invariant.
The nuclear vibration field of eigenvector $k$ is
\be
 u_k (\Oh) = \sum_{M} {1\over \sqrt{2}}
   (Y^*_{L_kM}(\Oh) b^\dagger_{L_kMk} + Y_{L_kM}(\Oh) b_{L_kMk})
\label{2.3}
\ee
where $\Oh$ is a unit vector on the sphere. The interaction between the
vibrations and electron density is
\be
H^{e-v}~\propto \sum_{k}  g_k\int\! d\Oh~ u_k(\Oh) \sum_s
\psi^\dagger_s(\Oh)\psi_s(\Oh)
\label{2.5}
\ee
where the electron field operators are
\be
\psi_s(\Oh)=\sum_{m=-1}^1 Y_{1m}(\Oh) c_{ms}
\label{2.4}
\ee
Using the relation
\be
\int \!d\Oh ~Y_{LM}(\Oh ) Y_{lm_1}(\Oh) Y_{lm_2}(\Oh) \propto (-1)^M \langle
L,-M|lm_1;lm_2\rangle , \label{2.6}
\ee
where $\langle \cdots \rangle$ is a Clebsch-Gordan coefficient \cite{ed},
yields the  second quantized Hamiltonian
\bea
H^{e-v}&=& H^{e-v}_{A_g} + H^{e-v}_{H_g} \nonumber\\
H^{e-v}_{A_g}&=& \sqrt{ 3\over 2 } \hbar  \sum_{k=1}^2 g_k\omega_k
\sum_{ms} (-1)^m \left(b^\dagger_{k0 }+ b_{k0}\right) \nonumber\\
&&~~~~~~~~~~~~~~~\times
\langle 0, 0|1,-m;1,m\rangle c^\dagger_{m s} c_{m s} \nonumber\\
H^{e-v}_{H_g}&=& { \sqrt{3}\over 2 } \hbar \sum_{k=3}^{10} g_k\omega_k
\sum_{M=-2}^{2} \sum_{ms} (-1)^m \left(b^\dagger_{kM }+(-1)^{M} b_{k~-M}\right)
 \nonumber\\
&&~~~~~~~~~~~~~~~\times
\langle 2, M|1,-m;1,m +M\rangle c^\dagger_{m s} c_{m+M  s} ~,
\label{2.7}
\eea
where the numerical constants are fixed by the requirement that the classical
JT energy gain of a single mode $k$ is $g_k^2/2$ \cite{I}.

\section{The Multi--Mode Spectrum (Weak Coupling)}
\label{MMS}

The perturbation hamiltonian (\ref{2.7}) written in the Fock basis
(\ref{basis}), connects states whose number of vibrons $N_{v}(k)$ of mode $k$
differs by exactly $\pm 1$. Thus, first order corrections to the energies
vanish. Second order corrections are obtained by diagonalizing the matrix
\cite{sak}
\be
\Delta^{(2)}_{a,b}=
 \langle a | H^{e-v} {1 \over {E^{(0)}_a-H^0}} H^{e-v} | b \rangle ,
\label{5.3}
\ee
where $ | a \rangle $ and $ | b \rangle $ are members of the same degenerate
manifold, i.e. they have the same number of vibrons $N_{v}(k)$. The sum implied
by the inverse operator $(E^{(0)}_a-H^0)^{-1}$ extends just to those states
whose $N_v$'s differ only by $\pm 1$ from that of the multiplet being
perturbed. This means that to second order in the coupling constants $g_k$
there are no direct inter-mode interactions, and the modes can be treated
separately. The only second-order inter-mode coupling is a consequence of all
the modes having a common ground state ($N_v$=0). The effect of vibron $k$
affects this $N_v$=0 state either with a pure shift or through
both shift and splitting. However
all other modes $k'$, having their ladder built on the same $N_v$=0 state, are
shifted or split by vibron $k$, according to the same structure of this $N_v$=0
multiplet. This effect takes place through additive contributions proportional
to $g_k^2$ to
the diagonal matrix elements $\Delta^{(2)}_{a,a}$ relative to the $k'$ ladder,
without involving off-diagonal inter-mode couplings, which would be related to
$g^4$ and higher order corrections.

A single $A_g$ mode coupled to a $t_{1u}$ level is the simple polaron problem,
which is exactly soluble \cite{mahan}: the second order energy is exact. Since
the $A_g$ representation is one-dimensional, it does not split the electronic
degeneracy. The only effect is a downward shift of the whole spectrum. The
amount of $A_g$-related energy shift is found to be $-E^{JT}_{ k}$, $-4E^{JT}_{
k}$ and $-6E^{JT}_{ k}$, for $n$=1, 2 and 3 respectively (where $E^{JT}_{ k}$
is the classical JT energy gain $\equiv g_k^2 \hbar \omega_k /2$ of that mode
$A_g(k)$). Obviously, these results hold for both unpolarized and polarized
spin states, since the $t_{1u}$ levels remain degenerate.

For the $H_g$ modes the situation is more complicated.
For $n$=1, the degenerate vibronic $t_{1u}$ ground state is not split by
JT coupling, as it conserves its $L$=1 symmetry. For this reason, the only
contribution of the vibron $H_g(k)$ to the spectrum of another vibron $H_g(k')$
is just a constant energy shift of -5/4 $g_k^2$,  which obviously does not
affect {\em energy differences} in the spectrum of vibron $k'$. This is
perfectly analogous to the effect of a single $A_g$ mode on all the other
vibrons.

By contrast, the $n$=2,4 and $n$=3 lowest vibronic multiplets ($N_v$=0) are
split (into $^1$S, $^1$D, $^3$P, and $^2$P, $^2$D, $^4$S levels respectively).
Correspondingly, the levels of an interacting vibron $k$ receive different
diagonal contributions of order $g_{k'}^2$ from different interacting vibrons
$k'$, giving rise to a more intricate pattern of splittings.

For the sake of simplicity, and also since C$_{60}^-$ seems easiest to obtain
in the gas phase \cite{yang}, we will concentrate on the many-modes spectrum
for $n$=1. The perturbative results for  all $H_g$ and $A_g$ modes will be
presented in Section \ref{spectros}. In principle, the  full spectra for
$n$=2,3,4, can be determined following the same method.

The spectrum for a single $H_g$
mode is given in Table I for the degenerate multiplets $N_v$=0 and
$N_v$=1, and for $n$=1, 2, ($S$=0) and 3 ($S$=1/2) electrons. In
Figures (\ref{ExactP1}, \ref{ExactP2}, \ref{ExactP3}) we replot on an expanded
scale the results of exact diagonalization of Ref \cite{I} for 1, 2 and 3
electrons, along with the straight lines corresponding to present perturbative
results for the lowest few states. These figures confirm that perturbative
results retain quantitative validity up to $g\approx 0.3$. In the special case
$n$=1, moreover, the perturbative results lay within 0.05$\hbar \omega$ of the
exact value up to $g\leq 0.4$.

We have therefore an approximate analytical estimate of the splittings induced
by JT coupling, valid in weak coupling. For example, for $n$=1 the $H_g$ vibron
excitation, originally at energy 1 above the ground state, splits into three
vibronic levels with relative shifts $-{3\over4}g^2$, ${3\over8}g^2$ and
${9\over8}g^2$ (in units of $\hbar \omega$ of that vibron). Table I contains
the complete list of these low lying excitations energies, accurate to order
$g^2$.

Because the effects of all $A_g$ and $H_g$ modes can be linearly superposed,
there are two ingredients only, which we need to have in order to transform the
analytical shifts of Table I into actual numbers for C$_{60}^{n-}$: the
frequencies $\hbar \omega_k$ and the coupling constants $g_k$ of each
individual $H_g$ and $A_g$ mode. For the frequencies, there are both calculated
and measured values. We can avoid uncertainties by choosing the latter as
given, e.g. for neutral C$_{60}$ in Ref.\cite{zhou}. In doing so, we neglect
the
well known small systematic frequency shifts associated with bond-length
readjustments and other electronic effects going from C$_{60}$ to C$_{60}^{n-}$
\cite{koh,modesti}. They also depend on the environment of the C$_{60}^{n-}$
ion.

There are several calculated sets of coupling constants $g_k$
\cite{negri,vzr,sch,antr}, but no direct measurement. Since the agreement among
the different calculations is far from good, in Section \ref{spectros} we
present the results of several selected sets,  which provides  an estimate of
the relevant uncertainties. We will eventually adopt the most recent values of
Antropov {\em et al}. \cite{antr}.

As  seen in Table II, almost all of the couplings $g_k$ are weak, $g_k \leq$
0.4. As discussed, in this range the perturbative results are accurate within
ten per cent or better for all the low lying states. As the discrepancies among
the various estimates of the $g_i$ is much larger, these perturbative formulas
are at this stage  more than adequate, and particularly good in the $n$=1 case,
where the distortion is smallest.

This is fortunate, since exact diagonalization is computationally rather
demanding if all $H_g$ modes are included. Better  knowledge of frequencies and
coupling constants might warrant a calculation of higher orders in $g_k$.

\section{Vibron spectroscopy of C$_{60}^{n-}$ anions.}
\label{spectros}

The electron affinity of C$_{60}$ is large (2.7 eV) and experimental evidence
has been found that the C$_{60}^-$ \cite{yang} and C$_{60}^{2-}$ \cite{limbach}
are stable ions in vacuum. In solution a wider spectrum of ionization states
has been demonstrated electrochemically, up to and including C$_{60}^{5-}$
\cite{dubois,heath,fullagar}. As an adsorbate on a metal surface, the
electronegative C$_{60}$ molecule naturally picks up electrons
\cite{erwin,burst}, and recent evidence has been provided of charge transfer
which can be as large as $n$=6 \cite{modesti}. In the solid state, finally,
there are compounds, covering a wide range of charge transfers, from $n$=1, as
in TDAE$^+$--C$_{60}^-$ \cite{denisov} or Rb$_1$C$_{60}$ \cite{benning}, $n$=3,
as in K$_3$C$_{60}$ or Rb$_3$C$_{60}$ \cite{hebard}, $n$=4 as in K$_4$C$_{60}$
\cite{kiefl}, $n$=6 as in Rb$_6$C$_{60}$ \cite{fisher}, or even higher as in
Li$_{12}$C$_{60}$ \cite{chabre,koh}.

Among these systems, our calculations so far address concretely only the gas
phase case. Unfortunately, to our knowledge no investigation appears to have
been made of the vibrational excitations of isolated C$_{60}^-$ and
C$_{60}^{2-}$. Our calculated excitation spectrum for C$_{60}^-$ therefore
constitutes a prediction which we hope will stimulate new work.

In Table II we report the excitation energies predicted by perturbation theory,
applied to  the eight modes in the C$_{60}^-$ case. Selection rules are not
discussed here for any particular spectroscopy. We simply give the symmetry
assignments.

As is seen, the predicted splittings due to dynamical JT coupling are generally
quite large, and should be well observed spectroscopically. However, as
indicated by comparison between different sets of $g_k$'s, there is a large
uncertainty in these predicted splittings of the same order of magnitude as the
splittings themselves. As remarked earlier, the same uncertainty does not
affect the energetics of the following section, which is on safer grounds. Our
calculated spectrum is therefore of qualitative value, and we rather expect it
to work backwards. That is, a future precise measurement of the splittings
should provide an accurate evaluation of the actual couplings.

As a further caution, we should stress that our spherical representations in
the Hamiltonian (\ref{2.7})  neglect interactions due to the icosahedral
lattice of carbons. For example the vibron multiplet of $L= 3$ decomposes due
to the lattice into $T_{2u} \oplus G_u$  \cite{ozaki} , etc. In addition to
neglecting lattice effects and  anharmonic interactions, we also ignore
spin-orbit coupling. As remarked in Part I, it has been shown \cite{gasyna} to
yield splittings of the order of 50 cm$^{-1}$ to the $L=$1 ground state (in Ar
matrix), which is not a negligible amount.  Thus we estimate that the
splittings obtained by our Hamiltonian should dominate the splittings found in
the real spectrum.

As pointed out by Bergomi and Jolicoeur \cite{bergomi}, experiments on anions
in matrix may be relevant to the vibronic effects. Near infrared and optical
spectra of C$_{60}^{n-}$ ions in solution are available \cite{heath}. A major
$t_{1u} \to t_{1g}$ optical transition near 1 eV is present for all $n$ values.
It is accompanied by additional vibronic shake-up structures, typically near
350, 750, 1400 and 1600 cm$^{-1}$. This limited information seems as yet
insufficient for any relevant comparison with our calculations. Well defined
vibrational spectra are instead available for chemisorbed C$_{60}^{n-}$
\cite{modesti} and for A$_n$C$_{60}$ alkali fullerides \cite{fisher}. In this
case, however, interaction of the electronic $t_{1u}$ level with surface states
or with other $t_{1u}$ states of neighbouring balls must turn the level into a
broad band, and our treatment as it stands is invalid. One can generally expect
rapid electron hopping from a molecule to another to interfere substantially
with the dynamical JT process, in a way which is not known at present. The
spectra of charged C$_{60}$ adsorbates and solids, in any case, do not present
evidence of any dynamic splittings such as those of Table II, but rather of a
gradual continuous shift most likely due to a gradual overall change of
geometry, as suggested also by LDA calculations \cite{andreoni}.

Summarizing, we are yet unaware of detailed spectroscopic confirmation of the
electron-vibron effects. We expect however these effects to be
observable in the gas phase of
C$_{60}^-$ and C$_{60}^{2-}$. In particular for  C$_{60}^-$ an observation that
the lowest $H_g$ vibron splits into a near 7-fold degenerate multiplet (i.e.
from the $L=3 \equiv T_{2u}
\oplus G_u$ pseudorotation level) would be an important confirmation of the
electron vibron theory and effects of Berry phases.

\section{Ground state energetics and effective Hubbard $U$'s.}
\label{gse}

Perturbation theory allows us to write analytic expressions for
the energy gain of the ground state at different $n$, and therefore for the
pair energy
\be
U_n= E_{n+1}+E_{n-1} -2E_n ~,
\label{U_n}
\ee
as discussed in Ref \cite{I}. Comparison with exact single-mode results shows a
systematical perturbative overestimate (Figures \ref{ExactP1}, \ref{ExactP2},
\ref{ExactP3}) of the $H_g$-related ground state energy shift. The error is
however relatively small and quite acceptable for the couplings in Table II.
The shift is 5/2 times larger than its classical value, as  discussed in Ref.
\cite{I}. This factor 5/2 is important, because it leads in turn to a
surprisingly large energetic lowering even for small $g$'s, making JT vibronic
coupling a much more important affair than it was understood so far. The
physical reason for the large energy gain is that the dynamically JT distorted
molecule undergoes a dramatic decrease of vibrational zero-point energy. This
adds an extra $-\frac{3}{4}g^2\hbar \omega$  (in the $n$=1 case, say) to the
{\em static} JT gain $-\half g^2\hbar \omega$ of each $H_g$ mode. The
zero-point energy decreases faster at small $g$ probably because the
mexican-hat potential well is more ``square-well''-like than the original
harmonic potential. We also note that the proliferation of excited states upon
coupling $H_g$ with $t_{1u}$ is of fermionic origin,
and does not add to the zero-point energy.

Within second  order perturbation theory
the ground state energy is a sum of all the 2+8 contributions of
the $A_g$ + $H_g$ modes:
\be
E_{tot}(n)=E_{A_g}+E_{H_g}= a_n E^{JT}_{A_g} + b_n E^{JT}_{H_g} =
a_n \sum_{k=1}^2 E^{JT}_{ k} + b_n \sum_{k=3}^{10} E^{JT}_{ k} ~,
\label{etot}
\ee
where
\be
	E^{JT}_{ k} \equiv  \half g_k^2 \hbar \omega_k ,
\ee
We already discussed in Section \ref{MMS} the JT ground state energy gains due
to an $A_g$ mode, with coefficients $a_n$=-1, -4, and -6, for $n$=1, 2 and 3
respectively. Table I gives the corresponding energies  for the
 $H_g(k)$ modes $(k>2)$. These are given by
\bea
	-\frac{5}{2}	E^{JT}_{2~k} & n=1 \nonumber\\
	-    10		E^{JT}_{2~k} & n=2 \nonumber\\
	-\frac{15}{2}	E^{JT}_{2~k} & n=3\label{EHg}
\eea
the appropriate coefficients being therefore $b_1$=-5/2, $b_2$=-10,
$b_3$=-15/2.

These expressions allow us to compute the individual contribution to the pair
energies $U_n$ (Eq. (\ref{U_n})) due to the $A_g$ and $H_g$ modes. We give
these formulae in Table III. The corresponding numerical values are reported in
Table IV, based on the ground state energy gains as given by the set of
coupling constants of Eq. (\ref{U_n}).

We consider the unpolarized spin sector. Similar
subtractions could easily be done, if needed, for high-spin states, or
high- and low-spin, using, for example $E_{tot}(n$=2,$S$=0), with
$E_{tot}(n$=3,$S$=3/2). Although, as we pointed out, the values of the
individual $g_k$'s of Ref. \cite{antr}, \cite{vzr} and \cite{sch} are
significantly uncertain, the global $E_{H_g}$ is much less author-dependent,
amounting to 102 meV, 84 meV and 78 meV respectively ($n$=1).

As Table II shows, the coupling with the $A_g$ mode pushes $U_1$ further
towards negative values, but has the opposite effect on $U_2$ and $U_3$.

The overall enhancement factor 5/2 in the ground state $H_g$ shift ends up
producing a much larger pair energy than expected so far based on classical JT
energies \cite{sch}. In particular, our calculated JT energy gain of $\approx$
0.4 eV for $n$=2 and $\approx$ 0.3 eV for $n$=3 (low-spin) is almost one order
of magnitude larger than the currently accepted values! This has important
implications, first of all, in determining whether the simple C$_{60}^{n-}$
ion, in vacuum, in a matrix or in solution, will choose to be high-spin or
low-spin.

In order to discuss this point, we recall the existence
of an intra-ball Coulomb repulsion ${\bf U}$ (not to be confused with the pair
energy of Ref \cite{I}), which for a $t_{1u}$ level is a matrix, specified
by two main values, $U_\parallel$ (two electrons in the same orbital),
and $U_\perp < U_\parallel$ (two electrons in different orbitals).
Using the JT energy differences between the high- and low-spin states as
\bea
E_{S=0}^{(2)} - E_{S=1}^{(2)} &=&
 (  U_\parallel^{(2)} - 4 E^{JT}_{A_g} - 10 E^{JT}_{H_g} )
  - (  U_\perp^{(2)} - 4 E^{JT}_{A_g} - \frac{5}{2} E^{JT}_{H_g})\nonumber\\
 &=&(  U_\parallel^{(2)} -  U_\perp^{(2)} ) - \frac{15}{2} E^{JT}_{H_g}
\approx  (  U_\parallel^{(2)} - U_\perp^{(2)} ) - 0.3 eV
\label{lowhi2} \\
E_{S=\half}^{(3)} - E_{S=\frac{3}{2}}^{(3)} &=&
(   U_\parallel^{(3)} + 2 U_\perp^{(3)}
- 6 E^{JT}_{A_g} - \frac{15}{2} E^{JT}_{H_g} )
- ( 3 U_\perp^{(3)} - 6 E^{JT}_{A_g} )\nonumber\\
 &=&(  U_\parallel^{(3)} -  U_\perp^{(3)} ) - \frac{15}{2} E^{JT}_{H_g}
\approx ( U_\parallel^{(3)} -  U_\perp^{(3)} ) - 0.3 eV ~,
\label{lowhi3}
\eea
where we have used the fact that the JT energetics for $n$=2, $S$=1 is
identical to that for $n$=1, $S$=1/2 \cite{I}, while for $n$=3, $S$=3/2 there
is no JT distortion. We have also used the $t_{1u}$ orbital unimodal and
bimodal splitting patterns of Part I to identify the filling ($n_1, n_2, n_3$).
In particular the fillings assumed are (0,0,2) for $n$=2, $S$=0; (1,1,0) for
$n$=2, $S$=1; (0,1,2) for $n$=3, $S$=1/2; (1,1,1) for $n$=3, $S$=3/2.

So long as $ {\bf U}^{(n)}$ may be expected to vary slowly with the electron
number $n$, then the two energy differences (\ref{lowhi2}) and (\ref{lowhi3})
should be very similar. Moreover, the prevailing of a high- or of a low-spin
state is decided by a fine balance between the Coulomb repulsion anisotropy $(
U_\parallel -  U_\perp )$ and the dynamical JT gain $\frac{15}{2} E^{JT}_{H_g}$
This suggests the possibility that if high-spin is more likely to prevail for
C$_{60}^{2-}$ and C$_{60}^{3-}$ in the gas phase, where ${\bf U}$ is large, the
balance might easily reverse in favor of low-spin when in matrix or in
solution. Recent EPR data indicate that this is precisely the case. When frozen
in a CH$_2$Cl$_2$ glass, C$_{60}^{2-}$ appears to be in a high-spin, $S$=1
state \cite{dubois}. Hence, in this case $( U_\parallel -  U_\perp )$  is
larger than 0.3 eV. However, optical and EPR data for C$_{60}^{3-}$ in
CH$_2$Cl$_2$ and other matrices favor a low-spin state \cite{heath}. Now $
U_\parallel^{(3)} - U_\perp^{(3)} $ has therefore become smaller than 0.3 eV.
We can conclude that, even for a single embedded molecule, the {\em balance
between intra-ball Coulomb repulsion and dynamical JT energy gains is extremely
critical}.

Recent photoemission and Auger data \cite{krum} have shown that the intra $h_u$
HOMO orbital Coulomb ${\bf U}$ is not as large as it was previously supposed.
In particular, a decrease by only 0.23 eV from gas phase \c60 and the
crystalline \c60 hole-hole Auger shifts, implies $|{\bf U}|<1$ eV in the
latter. This upper bound is about a factor three smaller than those previously
proposed \cite{lof}. In the light of this observation, it is not at all
surprising to find that $U_\parallel -  U_\perp$ is in the neighbourhood of 0.3
eV for C$_{60}^{n-}$ in a matrix.

Coming next to the pair binding energies of Ref \cite{I}, we find large
negative dynamical JT-related $U_n$'s for odd $n$. Even if we omit the $A_g$
contribution (which may be irrelevant for  superconductivity, due to screening
\cite{antr}), we get $U_3$=-0.2 eV. This negative value, will cancel at least a
good fraction of the Coulomb positive intra-ball pair energy
\be
  U_3^{Coul} = U_\parallel^{(2)} + (2 U_\parallel^{(4)} + 4 U_\perp^{(4)} ) -
2( U_\parallel^{(3)} + 2 U_\perp^{(3)} )
 \approx   U_\parallel
\label{UU3}
\ee
This cancellation implies a severe decrease of the Coulomb pseudopotential
$\mu^*$ relative to that calculated when the JT coupling is ignored
\cite{gunn}. For a sufficiently strong
solid-state screening of the electronic $U_\parallel$ and $ U_\perp$,
it may well be sufficient to reverse to a negative
$\mu^*$, i.e. to an overall negative Hubbard ${\bf U}$ state.

For $n$=2 and 4, dynamical JT stabilizes the average configuration of
$C_{60}^{n-}$, since the pair energy is {\em positive}: $U_2\approx 0.4$ eV.
This now acts to reinforce the bare Coulomb pair energy
\bea
  U_2^{Coul} = U_\parallel^{(3)}  - 2 U_\perp^{(3)} - 2 U_\parallel^{(2)}
 \approx  2 U_\perp - U_\parallel \nonumber\\
  U_4^{Coul} = (2 U_\parallel^{(5)} + 8 U_\perp^{(5)} )
	     + (  U_\parallel^{(3)} + 2 U_\perp^{(3)} )
	     -2(2 U_\parallel^{(4)} + 4 U_\perp^{(4)} )
 \approx  2 U_\perp - U_\parallel
\label{UU2}
\eea
where a filling (2,2,1) has been assumed for $n$=5.

For even $n$,  the JT coupling stabilizes a  {\em correlated insulating state}
of a lattice of evenly-charged  C$_{60}$ molecules. In this type of insulator,
fluctuations about $\langle n_i\rangle =n$ are suppressed, and a gap of order
$U_n$ is opened in the electronic spectrum.

This state has an even number of electrons per site, and is non-magnetic, very
much like a regular band insulator. However, electron correlations responsible
for band narrowing and gap opening are {\em vibronic} in origin. We suggest
that the (body-centered tetragonal \cite{fleming}) structure of K$_4$C$_{60}$
and Rb$_4$C$_{60}$ may be a realization of this state where electronic and
vibronic interactions play an important role. So far, band calculations
\cite{erwin} and experiments \cite{martin} had been in disagreement, the former
suggesting a metal, and the latter finding a narrow-gap insulator.

Very recent UPS data on K$_n$C$_{60}$ \cite{deseta} have shown a {\em decrease}
of the energy difference between the HOMO and the Fermi level (inside the
$t_{1u}$ LUMO) when going from $n$=3 to $n$=4 and finally to $n$=6. This kind
of non-rigid band behaviour is in itself not a surprising result. The surprise
is that the decrease is very large from $n$=3 to $n$=4 ($\approx$ 0.4 eV), and
smaller from $n$=4 to $n$=6 ($\approx$ 0.2 eV). As pointed out by De Seta and
Evangelisti, a positive Coulomb $U$ would predict exactly the opposite. We
observe that this behaviour is instead in agreement with our predicted pattern
of effective $U_n$ of vibronic origin, which is therefore supported by these
data.

Additional experiments which may probe the electron--vibron interactions are
short time resolved spectroscopy of excitons in neutral C$_{60}$
\cite{vardeny}. An exciton consists of an electron in the LUMO orbital and a
hole in the $H_g$  HOMO levels, which interact with different strengths with
the vibrons. The hole-vibron coupling inside the HOMO could be studied along
similar lines to those presented above for the $t_{1u}$ LUMO.

As for superconductivity in solids with $n$=3, we expect the enhanced pair
binding found here to be crucial for overcoming the on-site Coulomb repulsion
and for enhancing  $T_c$ over its value in, e.g., graphite intercalates.
Broadening of the $t_{1u}$ electron level into a band of non-negligible width
makes the present treatment insufficient for quantitative predictions. From the
fundamental point of view however, it is amusing to note that superconductivity
can be enhanced by a decrease of {\em lattice zero-point} energy. This adds to
the usual BCS mechanism of reducing the electron {\em kinetic} energy by
opening a gap. We hope to pursue this line of thought further in future work.

\section{Summary}
\label{Sec8}

In conclusion, a full treatment of all the $A_g$ and $H_g$ modes has been
given, and shown to yield analytical results with quantitative accuracy for the
full dynamical JT problem of C$_{60}^{n-}$. The ground state energetics has
been studied, and unexpectedly large energy gains have been found, due to a
decrease of zero-point energy. This implies large positive effective $U_n$ for
$n$=2 and 4, and a large negative $U_3$, which is very interesting in view of
superconductivity in K$_3$C$_{60}$ and insulating behaviour in K$_4$C$_{60}$.
Detailed vibrational spectra for C$_{60}^{n-}$ are presented, and proposed for
spectroscopic investigation, particularly in gas phase.

Related work is also being done by other groups \cite{others}.

\subsubsection*{Acknowledgements}

ET and NM wish to acknowledge discussions with W. Andreoni, E. Burstein, J.
Kohanoff, S. Modesti, L. Pietronero, P. Rudolf, C. Taliani and L. Yu. A.A.
thanks Mary O'Brien, Art Hebard and Zeev Vardeny for valuable discussions, and
acknowledges the Sloan  Foundation for a fellowship.  This paper was supported
by grants from the US-Israel Binational Science Foundation,  the Fund for
Promotion of Research at the Technion, and the US Department of Energy  No.
DE-FG02-91ER45441, the Italian Istituto Nazionale di Fisica della Materia INFM,
the European US Army Research Office, and NATO through CRG 920828.

\vfill\eject

\vfill\eject

\centerline{\underbar{\bf Table I}}
\vskip 0.3in
\begin{tabular}{|c|c|c|c|c|c|c|} \hline
   &     & original &$	2^{nd}$ order    	& residual &excit. energy&\\
$n$&$N_v$&degeneracy&shift ${\Delta^{(2)}}\over{g^2\hbar\omega}$&degeneracy&
${E^{(2)}-E^{(2)}_{ground}}\over{\hbar\omega}$ & $^{2S+1}\Re$\\ \hline
1 		& 0 &  3($\times$2)& -5/4 & 3& 0 		&$^2$P\\
($S=\frac{1}{2}$)&1 & 15($\times$2)& -2   & 7&$ 1-\frac{3}{4}g^2$	&$^2$F\\
		&   & 		   & -7/8 & 5&$ 1+\frac{3}{8}g^2$	&$^2$D\\
		&   & 		   & -1/8 & 3&$ 1+\frac{9}{8}g^2$	&$^2$P\\
\hline
2 		& 0 &  6	   & -5  & 1& 0  		 &$^1$S\\
($S=0$)		&   & 		   & -11/4& 5&$\frac{9}{4}g^2	$&$^1$D\\
  		& 1 & 30	   & -5  & 5&$ 1		$&$^1$D\\
		&   & 		   & -17/4& 9&$ 1+\frac{3}{4}g^2$&$^1$G\\
		&   & 		   & -11/4& 7&$ 1+\frac{9}{4}g^2$&$^1$F\\
		&   & 		   & -13/8& 5&$ 1+\frac{27}{8}g^2$&$^1$D\\
		&   & 		   & -7/8 & 3&$ 1+\frac{33}{8}g^2$&$^1$P\\
		&   & 		   & -1/2 & 1&$ 1+9g^2		$&$^1$S\\
\hline
2 		& 0 &  3($\times$3)& -5/4 & 3& 0		 &$^3$P\\
($S=1$)		& 1 & 15($\times$3)& -2   & 7&$ 1-\frac{3}{4}g^2$&$^3$F\\
		&   &		   & -7/8 & 5&$ 1+\frac{3}{8}g^2$&$^3$D\\
		&   &		   & -1/8 & 3&$ 1+\frac{9}{8}g^2$&$^3$P\\
\hline
3 		& 0 &  8($\times$2)& -15/4& 3& 0		 &$^2$P\\
($S=\frac{1}{2}$)&  & 		   & -9/4 & 5&$  \frac{3}{2}g^2	$&$^2$D\\
  		& 1 & 40($\times$2)& -9/2 & 7&$1-\frac{3}{4}g^2	$&$^2$F\\
		&   & 		   & -15/4& 9&$ 1		$&$^2$G\\
		&   & 		   & -15/4& 5&$ 1		$&$^2$D\\
		&   & 		   & -21/8& 3&$1+\frac{9}{8}g^2	$&$^2$P\\
		&   & 		   & -9/4 & 7&$1+\frac{3}{2}g^2	$&$^2$F\\
		&   & 		   & -9/8 & 5&$1+\frac{11}{8}g^2$&$^2$D\\
		&   & 		   & -3/8 & 3&$1+\frac{27}{8}g^2$&$^2$P\\
		&   & 		   &  0   & 1&$1+\frac{15}{4}g^2$&$^2$S\\
\hline
3 		& 0 &  1($\times$4)& 0	  & 1	& 0		 &$^4$S\\
($S=\frac{3}{2}$)&1 &  5($\times$4)& 0    & 5	& 1		 &$^4$D\\
\hline
\end{tabular}
\vskip 0.7in

TABLE I. Analytical expressions of energy shifts and excitation energies for
the
electron-vibron coupling of a single $H_g$ mode, for low-spin and
high-spin states, to second order in the coupling constant $g$.

\newpage

\vskip 3in
\centerline{\underbar{\bf Table II}}
\vskip 0.3in
\begin{tabular}{|c|c|c|c|l|} \hline
$H_g$&Exp.Energy & coupling	& excitation energy 		& \\
mode &(cm$^{-1}$)&	$g_k$	& $E_{fin}-E_{ground}$ 	 &$L_{fin}$ (sym.) \\
     &\cite{zhou}&\cite{antr} ~\cite{vzr} ~\cite{sch}&($cm^{-1}$)& \\ \hline
1 &270.0&0.33  0.33 0.54	& ~248 \ ~~248 \  ~212&3 ($T_{2u} \oplus G_u$)\\
  &	&              		& ~281 \ ~~281 \  ~299&2 \  ($H_u$)\\
  &	&              		& ~303 \ ~~303 \  ~357&1 \  ($T_{1u}$)\\
2 &430.5&0.37  0.15 0.40  	& ~387 \ ~~423 \  ~380&3 ($T_{2u} \oplus G_u$)\\
  &	&              		& ~452 \ ~~434 \  ~456&2 \  ($H_u$)\\
  &	&              		& ~496 \ ~~441 \  ~507&1 \  ($T_{1u}$)\\
3 &708.5&0.20  0.12 0.23  	& ~687 \ ~~701 \  ~679&3 ($T_{2u} \oplus G_u$)\\
  &	&              		& ~719 \ ~~712 \  ~723&2 \  ($H_u$)\\
  &	&              		& ~741 \ ~~719 \  ~752&1 \ ($T_{1u}$)\\
4 &772.5&0.19  0.00 0.30  	& ~751 \ ~~773 \  ~722&3 ($T_{2u} \oplus G_u$)\\
  &	&              		& ~783 \ ~~773 \  ~798&2 \ ($H_u$)\\
  &	&              		& ~805 \ ~~773 \  ~849&1 \ ($T_{1u}$)\\
5 &1099.0&0.16  0.23 0.09  	& 1077 \  1055 \  1092&3 ($T_{2u} \oplus G_u$)\\
  &	&              		& 1110 \  1121 \  1103&2 \ ($H_u$)\\
  &	&              		& 1132 \  1164 \  1110&1 \ ($T_{1u}$)\\
6 &1248.0&0.25  0.00 0.15  	& 1190 \  1248 \  1226&3 ($T_{2u} \oplus G_u$)\\
  &	&              		& 1277 \  1248 \  1259&2 \ ($H_u$)\\
  &	&              		& 1335 \  1248 \  1281&1 \ ($T_{1u}$)\\
7 &1426.0&0.37  0.48 0.30  	& 1281 \  1179 \  1332&3 ($T_{2u} \oplus G_u$)\\
  &	&              		& 1499 \  1549 \  1473&2 \ ($H_u$)\\
  &	&              		& 1644 \  1796 \  1568&1 \ ($T_{1u}$)\\
8 &1575.0&0.37  0.26 0.24  	& 1415 \  1495 \  1510&3 ($T_{2u} \oplus G_u$)\\
  &	&              		& 1655 \  1615 \  1608&2 \ ($H_u$)\\
  &	&		  	& 1815 \  1695 \  1673&1 \ ($T_{1u}$)\\
\hline
\end{tabular}
\vskip 0.7in

TABLE II. Vibronic excitation spectrum for the eight $H_g$ modes. Three
different sets of coupling constants used in the perturbative expressions of
Table I. The relations  \cite{sch} between the coupling strength $g$ and the
electron-phonon coupling $\lambda/N(\epsilon_F)$ (See  Ref. \cite{antr}) are
for $ H_g$ modes $g^2=\frac{6}{5}\lambda/N(\epsilon_F) / \hbar\omega$, and for
$A_g$ modes $ g^2=3\lambda/N(\epsilon_F) / \hbar\omega$.

\newpage

\centerline{\underbar{\bf Table III}}
\vskip 0.3in
\begin{tabular}{|c|c|c|c|} \hline
mode	&$U_1$		 &$U_2$			&$U_3$		\\ \hline
$A_g$	&$-2E^{JT}_{0~k}$&$E^{JT}_{0~k}$	&$4E^{JT}_{0~k}$	\\
$H_g$	&$-5E^{JT}_{2~k}$&$10E^{JT}_{2~k}$	&$-5E^{JT}_{2~k}$	\\
\hline
\end{tabular}
\vskip 0.7in

TABLE III. Analytical expressions for single mode pair energies (low-spin
states) to second order in the corresponding coupling constants $g_k$.

\newpage

\centerline{\underbar{\bf Table IV}}
\vskip 0.3in
\begin{tabular}{|c|c|c|rrr|} \hline
	&Exp.Energy 	& coupling	& \multicolumn{3}{|c|}{Ground state}\\
	&(cm$^{-1}$)	&	$g_k$	& \multicolumn{3}{|c|}{energy shift
(meV)} \\
     	&\cite{zhou}	&\cite{antr}	& $n$=1&$n$=2&$n$=3\\ \hline
$E_{A_g,1}$ &493.0	&0.38		& -5 & -18 &-27\\
$E_{A_g,2}$&1468.5	&0.39		& -14& -54 &-81\\
$E_{H_g,3}$ &270.0	&0.33		& -5 & -18 &-14\\
$E_{H_g,4}$ &430.5	&0.37		& -9 & -36 &-27\\
$E_{H_g,5}$ &708.5	&0.20		& -5 & -18 &-14\\
$E_{H_g,6}$ &772.5	&0.19		& -5 & -18 &-14\\
$E_{H_g,7}$&1099.0	&0.16		& -5 & -18 &-14\\
$E_{H_g,8}$&1248.0	&0.25		& -12& -48 &-36\\
$E_{H_g,9}$&1426.0	&0.37		& -30&-120 &-90\\
$E_{H_g,10}$&1575.0	&0.37		& -33&-132 &-99\\ \hline
$E_{A_g}$&		&		& -18& -72 &-108\\
$E_{H_g}$&		&		&-102&-408 &-306\\ \hline
$E_{tot}$ &		&		& -120&-480&-414\\ \hline
$U_{n,A_g}$&		&		& -36 &  18&  72\\
$U_{n,H_g}$&		&		&-204 & 408&-204\\
$U_{n,tot}$&		&		&-240 & 426&-132\\ \hline
\end{tabular}
\vskip 0.7in

TABLE IV. Dynamical JT ground state energy shifts due to each mode, their
total,
and the pair energies $U_n$. Results are accurate
to second order in the coupling constants $g_k$.

\newpage

\begin{figure}
\caption{Exact (solid line) and perturbative energies (dashed line) for $n$=1,
one coupled  $H_g$ mode of frequency $\omega$. The straight lines correspond
to second order perturbation theory for the $N_v$=0 and $N_v$=1 multiplets.}
\label{ExactP1}
\end{figure}

\begin{figure}
\caption{Exact (solid line) and perturbative energies (dashed line) for $n$=2,
one coupled  $H_g$ mode of frequency $\omega$. The perturbative lines
correspond  to  the $N_v$=0 and the lowest three multiplets of $N_v$=1.}
\label{ExactP2}
\end{figure}

\begin{figure}
\caption{Exact (solid line) and perturbative energies (dashed line) for $n$=2,
one coupled  $H_g$ mode of frequency $\omega$. The perturbative lines
correspond  to  the $N_v$=0 and the lowest three multiplets of $N_v$=1.  The
second order coupling does not split the $L$=4 and $L$=2 levels in the $N_v$=1
multiplet, while the exact theory finds they actually do separate. The next
excited level (of species P) is also drawn to show its crossing of the
initially lower D level}
\label{ExactP3}
\end{figure}


\begin{thebibliography}{99}
\bibitem{ehsymm} Because of electron-hole symmetry, fillings $n$ and 6-$n$ are
equivalent, leaving $n$=1, 2, 3 as the only inequivalent states.
\end{thebibliography}
\end{document}